\documentstyle[12pt]{article}
\begin{document}
\begin{flushright}
hep-th/0605213\\ CAS-BHU/Preprint
\end{flushright}
\vskip 1.5cm
\begin{center}
{\bf \Large {A generalization of the horizontality condition in
the superfield approach to nilpotent symmetries for QED with
complex scalar fields}}

\vskip 2cm

{\bf R.P.Malik}\footnote{ On leave of absence from S. N. Bose
National Centre for Basic Sciences, Block-JD, Sector-III, Salt
Lake, Kolkata-700 098, West Bengal, India. E-mail address:
malik@bose.res.in}\\ {\it Centre of Advanced Studies, Physics
Department,}\\ {\it Banaras Hindu University, Varanasi-221 005,
India} \\ {\bf E-mail address: malik@bhu.ac.in }\\

\vskip 2cm

\end{center}

\noindent {\bf Abstract}: We provide a generalization of the
horizontality condition of the usual superfield approach to
Becchi-Rouet-Stora-Tyutin (BRST) formalism to obtain the
(anti-)BRST symmetry transformations for {\it all} the fields of a
four (3 + 1)-dimensional interacting 1-form $U(1)$ gauge theory
(QED) within the framework of the augmented superfield formalism.
In the above interacting gauge theory, there is an explicit
coupling between the 1-form $U(1)$ gauge field and the complex
scalar fields. This interacting gauge field theory is considered
on the (4, 2)-dimensional supermanifold parametrized by the four
even spacetime variables $x^\mu$ (with $\mu = 0, 1, 2, 3)$ and a
pair of odd Grassmannian variables $\theta$ and $\bar\theta$. The
above nilpotent (anti-)BRST symmetry transformations are obtained
due to the imposition of a gauge (i.e. BRST) invariant restriction
on the appropriate superfields defined on the (4, 2)-dimensional
supermanifold. This restriction owes its origin to a pair of
(super) covariant derivatives and their intimate connection with
the 2-form (super) curvatures. The results obtained, due to the
application of the horizontality condition {\it alone}, are
contained in the results deduced due to the imposition of the
above gauge invariant restriction.\\

\baselineskip=16pt



\newpage

\noindent
{\bf 1 Introduction}\\

\noindent The usual superfield approach to
Becchi-Rouet-Stora-Tyutin (BRST) formalism [1-6] provides a deep
connection between some of the key mathematical properties
associated with the (anti-)BRST symmetries (as well as
corresponding generators) and the partial derivatives w.r.t. the
Grassmannian variables of the superspace coordinates that
characterize the $(D, 2)$-dimensional supermanifold on which a
given $D$-dimensional $p$-form ($p = 1, 2, 3......$) (non-)Abelian
gauge theory is considered. The above $(D, 2)$-dimensional
supermanifold is parametrized by the number $D$ of the commuting
spacetime variables $x^\mu$ (with $\mu = 0, 1, 2......D-1)$ and a
pair of anticommuting (i.e. $\theta^2 = \bar\theta^2 = 0, \theta
\bar\theta + \bar\theta \theta = 0$) Grassmannian variables
$\theta$ and $\bar\theta$. On this supermanifold, a $(p + 1)$-form
super curvature is constructed with the help of the super exterior
derivative $\tilde d = dx^\mu \partial_\mu + d \theta
\partial_\theta + d \bar\theta \partial_{\bar\theta}$ (with
$\tilde d^2 = 0$) and the super $p$-form gauge connection $\tilde
A^{(p)}$. This is subsequently equated, due to the well-known
horizontality condition [1-6], to the ordinary $(p + 1)$-form
curvature defined on the $D$-dimensional ordinary spacetime
manifold with the help of the ordinary exterior derivative $d =
dx^\mu \partial_\mu$ (with $d^2 = 0$) and the ordinary $p$-form
gauge connection $A^{(p)} = \frac{1}{p!} (dx^{\mu_1} \wedge
dx^{\mu_2} ......\wedge dx^{\mu_p}) A_{\mu_1 \mu_2.......\mu_p}$
that defines the $p$-rank antisymmetric tensor gauge potential of
the theory. The celebrated horizontality condition, christened as
the soul-flatness condition in [7], mathematically amounts to
setting equal to zero all the Grassmannian components of the $(p +
1)$-rank (anti)symmetric tensor that constitutes the $(p +
1)$-form super curvature defined on the above supermanifold.

The above horizontality condition has been extensively exploited
to derive the nilpotent (anti-)BRST symmetry transformations for
the gauge field and corresponding fermionic (anti-)ghost fields of
the four $(3 + 1)$-dimensional (4D) 1-form (i.e. $A^{(1)} = dx^\mu
A_\mu $) non-Abelian gauge theory. To be more specific, this 4D
non-Abelian theory is first considered on a (4, 2)-dimensional
supermanifold parametrized by the four spacetime (even)
coordinates $x^\mu (\mu = 0, 1, 2, 3)$ and a pair of Grassmannian
(odd) variables $\theta$ and $\bar \theta$. An appropriate
$2$-form super curvature $\tilde F^{(2)} = \tilde d \tilde A^{(1)}
+ \tilde A^{(1)} \wedge \tilde A^{(1)}$ is constructed with the
help of the super exterior derivative $\tilde d$ and  super
$1$-form connection $\tilde A^{(1)}$. This is equated to the
ordinary 2-form curvature $F^{(2)} = d A^{(1)} + A^{(1)} \wedge
A^{(1)}$ constructed with the help of the ordinary exterior
derivative $d = dx^\mu \partial_\mu$ and ordinary 4D 1-form
connection $A^{(1)}$. This equality (i) leads to the derivation of
nilpotent (anti-)BRST symmetry transformations, and (ii) provides
the geometrical interpretation for the nilpotent (anti-)BRST
charges (and the nilpotent symmetry transformations they generate)
in the language of the translational generators along the
Grassmannian directions of the (4, 2)-dimensional supermanifold.
However, the latter type of geometrical interpretations remain
confined {\it only} to the gauge field and the corresponding
(anti-)ghost fields of the theory. The matter fields of the
interacting 1-form non-Abelian gauge theory remain untouched in
the framework of the usual superfield formulation (with the
theoretical arsenal of horizontality condition alone).

The above trick has also been exploited in the context of the
derivation of the nilpotent (anti-)BRST symmetry transformations
for the 2-form (i.e. $A^{(2)} = \frac{1}{2!} (dx^\mu \wedge
dx^\nu) B_{\mu\nu}$) non-interacting Abelian gauge theory in 4D
where a $3$-form super curvature $\tilde F^{(3)} = \tilde d \tilde
A^{(2)}$, constructed with the help of the super exterior
derivative $\tilde d$ and super 2-form connection $\tilde
A^{(2)}$, is equated to the ordinary 4D  $3$-form $F^{(3)} = d
A^{(2)}$ constructed with the help of the ordinary exterior
derivative $d$ and 2-form ordinary connection $A^{(2)}$. As
expected, here too, one obtains the nilpotent (anti-)BRST
transformations for the second rank antisymmetric  gauge field
$B_{\mu\nu}$ and the corresponding (anti)commuting (anti-)ghost
fields of the theory. Of course, the (anti-)ghost fields turn out
to be bosonic as well as fermionic in nature for the 2-form
Abelian gauge theory. Even for this Abelian theory, the matter
fields are not discussed at all in the framework of the usual
superfield approach to BRST formalism.

In a recent set of papers [8-17], the above horizontality
condition of the usual superfield approach has been consistently
extended so as to derive the nilpotent (anti-)BRST symmetry
transformations for the matter fields together with the above
nilpotent transformations associated with the gauge and
(anti-)ghost fields. This extended version of the superfield
formalism has been christened as the augmented superfield
formalism where, in addition to the horizontality condition, some
other physically interesting restrictions are imposed on the
superfields of the appropriately chosen supermanifold. In the
latter category of restrictions, mention can be made of the
equality of (i) the conserved currents corresponding to the gauge
symmetries [8-10,13], (ii) any conserved quantities for the
reparametrization invariant theories [11,12], and (iii) the gauge
invariant quantities owing their origin to the covariant
derivatives on the appropriately chosen superfields [14-17]. One
obtains logically consistent nilpotent (anti-)BRST symmetry
transformations for all the fields due to the application of the
restrictions (i) and (ii). However, the application of the
restriction (iii) on the superfields (defined on the appropriately
chosen supermanifolds) leads to the derivation of mathematically
exact nilpotent (anti-)BRST symmetry transformations for all the
fields. In a very recent set of papers [18,19], the nilpotent
(anti-)BRST symmetry transformations for all the fields of the
interacting 4D (non-)Abelian gauge theories (with the Dirac fields
as the interacting matter fields) have been derived from a {\it
single} restriction on the appropriate superfields of the
supermanifold. These attempts have been made to generalize the
horizontality condition to obtain all the nilpotent (anti-)BRST
symmetry transformations for all the fields of a given gauge
theory without spoiling the geometrical interpretations of some of
the key properties associated with the nilpotent (anti-)BRST
symmetries (and corresponding nilpotent charges) that are provided
by the horizontality condition alone. All the above mathematically
consistent extensions of the usual superfield formalism are called
by us as the augmented superfield approach to BRST formalism.

The central theme of our present paper is to demonstrate  that the
ideas of the augmented superfield formalism, with a single gauge
(i.e. BRST) invariant restriction on the appropriately chosen
superfields (defined on a suitable supermanifold) [18,19], can be
extended to derive the on-shell as well as off-shell nilpotent
(anti-)BRST symmetry transformations for {\it all} the fields of
an interacting four $(3 + 1)$-dimensional $U(1)$ gauge theory
where there is an explicit coupling between the $U(1)$ gauge field
and the charged complex scalar fields. We show that all the
results, obtained due to the application of the horizontality
condition of the usual superfield formulation, are contained in
the results deduced by exploiting our present gauge (i.e. BRST)
invariant restriction on the appropriately chosen superfields. On
top of it, the appropriate modifications of our present
restriction on the superfields (defined on a suitably chosen
supermanifold) provides a precise way to derive the on-shell
nilpotent (anti-)BRST transformations for all the fields
(including the matter fields) of the theory in a separate and
independent manner. It should be re-emphasized that the
horizontality condition {\it alone} does not shed any light on the
derivation of the nilpotent symmetry transformations associated
with the matter fields of any arbitrary interacting gauge theory
in any arbitrary dimension of spacetime. Thus, our present
endeavour is an important  step in the direction to generalize the
horizontality condition of the usual superfield approach to a more
general condition on the appropriately chosen superfields. Our
present Abelian gauge (i.e. BRST) invariant restriction owes its
origin to a pair of (super) covariant derivatives, their operation
on matter (super) fields and their intimate connection with the
Abelian (super) curvature 2-forms defined on the appropriately
chosen (super) spacetime manifolds. In our present investigation,
as a warm up exercise, we first derive the on-shell nilpotent BRST
and anti-BRST symmetry transformations for all the fields of the
present interacting U(1) gauge theory by invoking the chiral and
anti-chiral superfields defined on the (4, 1)-dimensional super
sub-manifolds (cf. section 3 below). Later on, we merge together
these superfields to obtain the general (4, 2)-dimensional
superfields for the derivation of the off-shell nilpotent
(anti-)BRST symmetry transformations {\it together} for all the
fields of our present gauge theory from a single restriction on
the matter superfields (cf. section 4 below). Thus, our main
results are contained in sections 3 and 4.

The compelling reasons behind our present investigation are
primarily four in number. First and foremost, it is important to
{\it generalize} our ideas of the augmented superfield approach to
BRST formalism [18,19] to a new field theoretical system where a
{\it single} gauge (i.e. BRST) invariant restriction on the
superfields (defined on the appropriately chosen supermanifolds)
leads to the derivation of the nilpotent (anti-)BRST symmetry for
{\it all} the fields of the interacting gauge theory. In fact, the
above ideas have been found to be true for the derivation of the
nilpotent symmetry transformations in the cases of (i) an
interacting 4D Abelian U(1) gauge theory where there is an
explicit coupling between the gauge field and the Dirac (matter)
fields, and (ii) an interacting 4D non-Abelian gauge theory where
the Noether conserved current, constructed with the help of Dirac
fields, couples to the $SU(N)$ non-Abelian gauge field. In our
present endeavour, we exploit primarily the same restriction for a
new field theoretic model where there is a coupling between the
$U(1)$ gauge field and the Noether conserved current constructed
with the help of the charged complex scalar fields and the gauge
field itself. Thus, our present investigation is essential to put
our earlier ideas on a firmer ground. Second, in our earlier works
[13,14], connected with the superfield approach to the derivation
of nilpotent symmetry transformations for the complex scalar
fields, we exploited two separate restrictions on the (4,
2)-dimensional supermanifold which included the horizontality
condition of the usual superfield formalism as one of the
restrictions. Our present endeavour is more economical and
aesthetically more appealing because we derive all the nilpotent
(anti-)BRST symmetry transformations from a {\it single} gauge
invariant restriction. Third, the single restriction exploited in
our present investigation is physically more appealing because
this restriction is a gauge (i.e. BRST) invariant condition on the
suitably chosen superfields. In contrast, the horizontality
condition is intrinsically a gauge covariant restriction on the
gauge superfield. Finally, our present field theoretical model is
(i) useful in the context of gauge theory of standard model, and
(ii) different from the interacting 4D (non-)Abelian gauge
theories with fermionic Dirac fields. For instance, the present
model allows an inclusion of a gauge (i.e. BRST) invariant
potential with a quartic renormalizable self-interaction term.
Such kind of interaction term is forbidden for the fermionic Dirac
fields in interaction with the (non-)Abelian gauge fields.

The contents of our present paper are organized as follows. In
section 2, we recapitulate the key points of the nilpotent
(anti-)BRST symmetry transformations for all the fields of an
interacting $U(1)$ gauge theory in the Lagrangian formulation
where there is an explicit coupling between the $U(1)$ gauge field
and the complex scalar fields. Section 3 is devoted to the
derivation of the on-shell nilpotent (anti-)BRST symmetry
transformations for the appropriate fields of the above
interacting gauge theory in the framework of the superfield
formulation. In this derivation, the (anti-)chiral superfields are
invoked for the gauge invariant restriction. The above specific
superfields are defined on the (4, 1)-dimensional (anti-)chiral
super sub-manifold of the general (4, 2)-dimensional
supermanifold. The material of section 4 deals with the derivation
of the off-shell nilpotent (anti-)BRST symmetry transformations
for all the fields of the theory by exploiting a gauge invariant
restriction on the general superfields of the (4, 2)-dimensional
supermanifold. An alternative version of this section is presented
in the Appendix A. Finally, in section 5, we make some concluding
remarks and point out a few future directions for further
investigations.\\

\noindent {\bf 2 (Anti-)BRST symmetries: Lagrangian formalism}\\

\noindent To provide a brief synopsis of the salient features of
the off-shell as well as on-shell nilpotent (anti-)BRST
symmetries, we focus on the Lagrangian density of an {\it
interacting} four ($3 + 1)$-dimensional \footnote{We adopt here
the conventions and notations such that the 4D flat Minkowski
metric is: $\eta_{\mu\nu} =$ diag $(+1, -1, -1, -1)$ and $\Box =
\eta^{\mu\nu} \partial_{\mu} \partial_{\nu} = (\partial_{0})^2 -
(\partial_{i})^2, F_{0i} = \partial_{0} A_{i} -
\partial_{i} A_{0} = E_i \equiv {\bf E}, F_{ij} = \epsilon_{ijk}
B_k, B_i \equiv {\bf B} = \frac{1}{2} \epsilon_{ijk} F_{jk},
(\partial \cdot  A) =
\partial_{0} A_0 - \partial_i A_i$ where ${\bf E}$ and ${\bf B}$
are the electric and magnetic fields, respectively and
$\epsilon_{ijk}$ is the totally antisymmetric Levi-Civita tensor
defined on the 3D (space) sub-manifold of the 4D spacetime
manifold. Here the Greek indices: $\mu, \nu, \lambda... = 0, 1, 2,
3$ correspond to the spacetime directions and Latin indices $i, j,
k, ...= 1, 2, 3$ stand only for the space directions on the
Minkowski spacetime manifold.} (4D) $U(1)$ gauge theory which
describes a dynamically closed system of the charged complex
scalar fields and U(1) gauge field. The (anti-)BRST invariant
Lagrangian density of the above system, in the Feynman gauge, is
[7,20-22] $$
\begin{array}{lcl}
{\cal L}_{b} &=& - \frac{1}{4}\; F^{\mu\nu} F_{\mu\nu} + \bar
D_\mu \phi^{*} D^\mu \phi - V (\phi^* \phi) + B \;(\partial \cdot
A) + \frac{1}{2}\; B^2 - i \;\partial_{\mu} \bar C \partial^\mu C
\nonumber\\ &\equiv& \frac{1}{2}\; ({\bf E^2} - {\bf B^2}) + \bar
D_\mu \phi^{*} D^\mu \phi - V (\phi^* \phi) + B \;(\partial \cdot
A) + \frac{1}{2}\; B^2 - i \;\partial_{\mu} \bar C
\partial^\mu C
\end{array} \eqno(2.1)
$$ where $V(\phi^*\phi)$ is the potential describing the quadratic
and quartic  interactions between the complex scalar fields $\phi$
and $\phi^*$ \footnote{This potential can be chosen in the quartic
polynomial form as: $V (\phi^* \phi) = \alpha^2 \phi^* \phi +
\beta (\phi^* \phi)^2$ for a renormalizable quantum field theory.
Here $\alpha$ and $\beta$ are the parameters which could be chosen
in different ways for different purposes (see, e.g. [23]). The key
point to be noted is the fact that this potential remains
invariant under the $U(1)$ gauge transformations as well as the
(anti-)BRST symmetry transformations.} . The covariant derivatives
on these fields, that are endowed with the electric charge $e$,
are as given below $$
\begin{array}{lcl}
D_\mu \phi = \partial_\mu \phi + i e A_\mu \phi\;\; \qquad\;\;
\bar D_\mu \phi^* = \partial_\mu \phi^* - i e A_\mu \phi^*.
\end{array} \eqno(2.2)
$$ In the Lagrangian density (2.1), the Nakanishi-Lautrup
auxiliary field $B$ is required to linearize the gauge-fixing term
$-\frac{1}{2} (\partial \cdot A)^2$. The Faddeev-Popov
(anti-)ghost fields $(\bar C)C$ (with $C^2 = \bar C^2 = 0, C \bar
C + \bar C C = 0$) are needed in the theory to maintain the
``quantum'' gauge (i.e. BRST) invariance and unitarity {\it
together} at any arbitrary order of perturbative calculations
(see, e.g. [24]). In the sense of the basic requirements of a
canonical field theory, the Lagrangian density ${\cal L}_b$
describes a dynamically closed system of all the fields $\phi,
\phi^*, C, \bar C$ and $A_\mu$ (see, e.g., [23]). It will be noted
that the gauge field $A_\mu$ couples to the conserved matter
current $J^{(c)}_\mu \sim (\phi^* D_\mu \phi - \phi \bar D_\mu
\phi^*)$ to provide the interaction between $A_\mu$ and matter
fields $\phi$ and $\phi^*$. This statement can be succinctly
expressed as $$
\begin{array}{lcl}
{\cal L}_{b} &=& - \frac{1}{4}\; F^{\mu\nu} F_{\mu\nu} +
\partial_\mu \phi^{*} \partial^\mu \phi - i e A_\mu [\phi^*
\partial_\mu \phi - \phi \partial_\mu \phi^*] + e^2 A^2 \phi^*
\phi\nonumber\\ &-& V (\phi^* \phi) + B \;(\partial \cdot A) +
\frac{1}{2}\; B^2 - i \;\partial_{\mu} \bar C \partial^\mu C.
\end{array} \eqno(2.3)
$$ The conservation of the matter current $J^{(c)}_\mu$ can be
easily checked by exploiting the equations of motion $D_\mu D^\mu
\phi = - (\partial V/\partial \phi^*), \bar D_\mu \bar D^\mu
\phi^*  = - (\partial V/\partial \phi)$ derived from the above
Lagrangian densities. These Lagrangian densities respect the
following off-shell nilpotent ($s_{(a)b}^2 = 0$) and anticommuting
($s_b s_{ab} + s_{ab} s_b = 0$) (anti-)BRST symmetry
transformations $s_{(a)b}$ \footnote{We follow here the notations
and conventions adopted in [21,22]. In fact, the (anti-)BRST
prescription is to replace the local gauge parameter by an
anticommuting number $\eta$ and the (anti-)ghost fields $(\bar
C)C$ which anticommute (i.e. $\eta C + C \eta = 0, \eta \bar C +
\bar C \eta = 0$) and commute with all the fermionic $(i.e. C \bar
C + \bar C C = 0, C^2 = \bar C^2 = 0$, etc.) and bosonic fields,
respectively. In its totality, the  nilpotent ($ \delta_{(A)B}^2 =
0$) (anti-)BRST transformations $\delta_{(A)B}$ are the product
(i.e. $\delta_{(A)B} = \eta s_{(a)b}$) of $\eta$ and $s_{(a)b}$
where $s_{(a)b}^2 = 0$.} for the matter fields, gauge field and
the (anti-)ghost fields, namely; $$
\begin{array}{lcl}
s_{b} A_{\mu} &=& \partial_{\mu} C\;\; \qquad \;s_{b} C = 0\;\;\;
\qquad s_{b} \bar C = i B\;\;\;  \qquad\; s_b \phi = - i e C \phi
\nonumber\\ s_b  \phi^* &=& + i e  \phi^* C\;\; \qquad s_{b} {\bf
B} = 0\;\; \quad  s_{b} B = 0\; \quad \;s_{b} {\bf E} = 0\; \quad
s_b (\partial \cdot A) = \Box C \nonumber\\ s_{ab} A_{\mu} &=&
\partial_{\mu} \bar C\;\; \qquad s_{ab} \bar C = 0\; \qquad\; s_{ab}
C = - i B\;\;  \qquad\; s_{ab} \phi = - i e \bar C \phi
\nonumber\\ s_{ab} \phi^* &=& + i e  \phi^* \bar C\; \qquad s_{ab}
{\bf B} = 0\; \quad  s_{ab} B = 0\; \quad \;s_{ab} {\bf E} = 0\;
\quad s_{ab} (\partial \cdot A) = \Box \bar C.
\end{array}\eqno(2.4)
$$ The key points to be noted, at this stage, are (i) under the
(anti-)BRST transformations, it is the kinetic energy term $(-
\frac{1}{4} F^{\mu\nu} F_{\mu\nu})$ of the gauge field $A_\mu$
which remains invariant. This statement is true for 1-form
(non-)Abelian gauge theories. For the above $U(1)$ gauge theory,
as it turns out, it is the curvature term $F_{\mu\nu}$ itself that
remains invariant under the (anti-)BRST transformations. (ii) In
the mathematical language, the (anti-)BRST symmetry
transformations owe their origin to the exterior derivative $d =
dx^\mu \partial_\mu$ because the curvature term, owing its origin
to the 2-form $F^{(2)} = d A^{(1)}$, is constructed from it and
the 1-form connection $A^{(1)} = dx^\mu A_\mu$. (iii) One can
obtain the on-shell ($\Box C = \Box \bar C = 0$) nilpotent
($\tilde s_{(a)b}^2 = 0$) (anti-)BRST symmetry transformations
$\tilde s_{(a)b}$ for the above theory from (2.4), by the
substitution $B = - (\partial \cdot A)$, as given below $$
\begin{array}{lcl}
\tilde s_{b} A_{\mu} &=& \partial_{\mu} C\;\; \qquad \tilde s_{b}
C = 0\;\; \qquad \tilde s_{b} \bar C = - i (\partial \cdot A)\;\;
\qquad \tilde s_b \phi = - i e C \phi \nonumber\\ \tilde s_b
\phi^* &=& + i e \phi^* C\;\; \qquad \;\tilde s_{b} {\bf E} =
0\;\; \qquad\; \tilde s_b {\bf B} = 0\; \qquad \;\tilde s_b
(\partial \cdot A) = \Box C \nonumber\\ \tilde s_{ab} A_{\mu} &=&
\partial_{\mu} \bar C\;\; \qquad \tilde s_{ab} \bar C = 0\;\; \qquad
\tilde s_{ab} C = + i (\partial \cdot A)\;\;  \qquad \tilde s_{ab}
\phi = - i e \bar C \phi \nonumber\\ \tilde s_{ab} \phi^* &=& + i
e \phi^* \bar C\;\; \qquad \;\tilde s_{ab} {\bf B} = 0\; \qquad
\;\tilde s_{ab} {\bf E} = 0\;\; \qquad \tilde s_{ab} (\partial
\cdot A) = \Box \bar C.
\end{array}\eqno(2.5)$$
The above local, infinitesimal, anticommuting and on-shell
nilpotent transformations are the symmetry transformations for the
following Lagrangian density $$
\begin{array}{lcl}
\tilde {\cal L}_{b} &=& - \frac{1}{4}\; F^{\mu\nu} F_{\mu\nu} +
\partial_\mu \phi^{*} \partial^\mu \phi - i e A_\mu [\phi^*
\partial_\mu \phi - \phi \partial_\mu \phi^*] + e^2 A^2 \phi^*
\phi\nonumber\\ &-& V (\phi^* \phi) - \frac{1}{2}(\partial \cdot
A)^2 - i \;\partial_{\mu} \bar C \partial^\mu C
\end{array} \eqno(2.6)
$$ which is derived from (2.3) by the substitution $B = -
(\partial \cdot A)$. (iv) In general, the above transformations
can be concisely expressed in terms of the generic fields $\Omega
(x), \tilde \Omega (x) $ and the conserved charges $Q_{(a)b},
\tilde Q_{(a)b}$, as $$
\begin{array}{lcl}
s_{r}\; \Omega (x) = - i\; \bigl [\; \Omega (x),  Q_r\; \bigr
]_{\pm}\; \qquad \tilde s_r \tilde \Omega (x) = - i \bigl [\;
\tilde \Omega (x), \tilde Q_r \bigr ]_{\pm}\; \qquad  r = b, ab
\end{array} \eqno(2.7)
$$ where the local generic fields $\Omega = A_\mu, C, \bar C,
B,\phi, \phi^*$ and $ \tilde \Omega = A_\mu, C, \bar C, \phi,
\phi^*$ are the fields of the Lagrangian densities (2.3) and
(2.6). The $(+)-$ signs, as the subscripts on the square bracket
$[\;, \;]_{\pm}$, stand for the bracket to be an (anti)commutator
for $\Omega, \tilde \Omega$ being (fermionic)bosonic in nature.
The explicit forms of the conserved, anticommuting  and nilpotent
(anti-)BRST charges $Q_r, \tilde Q_r, (r = b , ab)$ are not
required for our present discussion but can be derived for the
symmetry transformations (2.4) and (2.5)(Noether theorem).\\

\noindent {\bf 3 On-shell nilpotent symmetries: superfield
formalism}\\

\noindent In this section, we first focus on the derivation of the
on-shell nilpotent BRST symmetry transformations for {\it all} the
fields and, later on, we derive the anti-BRST symmetry
transformations for {\it all} the fields by invoking the potential
and power of specific restrictions on the chiral and anti-chiral
superfields (defined on the (4, 1)-dimensional super sub-manifolds
of the general (4, 2)-dimensional supermanifold), respectively.\\

\noindent
 {\bf 3.1 On-shell nilpotent BRST symmetries: chiral
 superfields}\\

\noindent To obtain the on-shell nilpotent BRST symmetry $\tilde
s_b$  transformations (2.5) for the basic fields of the Lagrangian
density (2.6), first of all, we generalize the 4D basic fields
$A_\mu, C, \bar C, \phi, \phi^*$ to the corresponding chiral
($\theta = 0$) superfields defined on the (4, 1)-dimensional super
sub-manifold of the general (4, 2)-dimensional supermanifold.
These chiral superfields can be expanded in terms of the basic
fields and some secondary fields (e.g. $R_\mu, B_1, B_2, f_1,
f_2^*$) as $$
\begin{array}{lcl}
{\cal B}^{(c)}_\mu (x, \bar\theta) &=& A_\mu (x) + \bar\theta\;
R_\mu (x)\; \qquad {\cal F}^{(c)} (x, \bar\theta) = C (x) + i \;
\bar\theta\; B_1 (x) \nonumber\\ \bar {\cal F}^{(c)} (x,
\bar\theta) & = &  C (x) + i \; \bar\theta\; B_2 (x)\; \qquad
\Phi_{(c)} (x, \bar\theta)  =  \phi (x) + i \; \bar\theta\; f_1
(x) \nonumber\\ \Phi^*_{(c)} (x, \bar\theta) &=& \phi^* (x) + i \;
\bar\theta\; f^*_2 (x).
\end{array}\eqno(3.1)
$$ The noteworthy points, at this stage, are:\\

\noindent (i) The chiral superfields ${\cal B}_\mu, \Phi, \Phi^*$
are bosonic (i.e. $({\cal B}_\mu)^2 \neq 0, (\Phi)^2 \neq 0,
(\Phi^*)^2 \neq 0$) in nature whereas the superfields ${\cal F},
\bar {\cal F}$ are fermionic [i.e. ${\cal F}^2 = 0, (\bar {\cal
F})^2 = 0, {\cal F} \bar {\cal F} + \bar {\cal F} {\cal F} = 0$].

\noindent (ii) In the limit $\bar\theta \to 0$, one retrieves the
basic fields of the Lagrangian density (2.6).

\noindent (iii) The number of fermionic fields $C, \bar C, f_1,
f_2^*, R_\mu$ do match with the number of bosonic fields $B_1,
B_2, \phi, \phi^*, A_\mu$ on the right hand side of the above
super expansions.

\noindent (iv) All the fields, on the r.h.s. of the above
expansion, are function of the 4D coordinates $x^\mu$ only because
they have been expanded along the $\bar\theta$-direction of the
super sub-manifold.

The following gauge (i.e. BRST) invariant restriction \footnote{It
will be noted that there exists another gauge (i.e. the on-shell
nilpotent BRST) invariant restriction on the chiral superfields
defined on the (4, 1)-dimensional chiral super sub-manifold (of
the general (4, 2)-dimensional supermanifold) that also leads to
the derivation of the on-shell nilpotent BRST transformations
$\tilde s_b$ of (2.5). This restriction is: $\Phi_{c} (x,
\bar\theta) \tilde {\bar {\cal D}}_{(c)} \tilde {\bar {\cal
D}}_{(c)} \Phi^*_{(c)} (x, \bar\theta) = \phi (x) \bar D \bar D
\phi^* (x)$ where $\bar D \phi^* = dx^\mu (\partial_\mu - i e
A_\mu) \; \phi^*$ and $\tilde {\bar {\cal D}}_{(c)} = dx^\mu
(\partial_\mu - i e {\cal B}^{(c)}_\mu) + d \bar\theta
(\partial_{\bar\theta} - i e {\cal F}^{(c)})$. It is evident that
the r.h.s. of this restriction is: $- i e \phi (x) F^{(2)} \phi^*
(x)$. This is a gauge (i.e. BRST) invariant quantity because
$\tilde s_b (\phi F_{\mu\nu} \phi^*) = 0$ as can be seen by
exploiting the on-shell nilpotent BRST symmetry transformations
(2.5).} on the chiral superfields $\phi_{(c)} (x, \bar\theta)$ and
$\phi^*_{(c)} (x, \bar\theta)$ (defined on the (4, 1)-dimensional
chiral super sub-manifold of the general (4, 2)-dimensional
supermanifold), namely; $$
\begin{array}{lcl}
\Phi^*_{(c)} (x, \bar\theta)\; {\cal D}_{(c)}\; {\cal D}_{(c)}
\Phi_{(c)} (x, \bar\theta) = \phi^* (x) \;D\; D\; \phi (x)
\end{array}\eqno(3.2)
$$ leads to the derivation of all the on-shell nilpotent BRST
transformations $\tilde s_b$ quoted in (2.5). In the above, the
covariant derivative $D \phi (x) = (d + i e A^{(1)}) \phi(x)
\equiv dx^\mu (\partial_\mu + i e A_\mu) \phi (x)$ where the
exterior derivative $d = dx^\mu \partial_\mu$ and 1-form
connection $A^{(1)} = dx^\mu A_\mu$. These quantities, defined on
the ordinary 4D spacetime manifold, are generalized to the (4,
1)-dimensional chiral super sub-manifold of the general (4,
2)-dimensional supermanifold, as  $$
\begin{array}{lcl}
{\cal D}_{(c)} &=& (\tilde d_{(c)} + i e \tilde A^{(1)}_{(c)})
\equiv  d x^\mu\; (\partial_\mu + i e {\cal B}_\mu^{(c)}) + d
\bar\theta \;(\partial_{\bar\theta} + i e {\cal F}^{(c)})
\nonumber\\ \tilde d_{(c)} &=& dx^\mu
\partial_\mu + d \bar\theta \partial_{\bar\theta}
\equiv d Z^M_{(c)} \partial_M^{(c)} \qquad \tilde A^{(1)}_{(c)} =
d Z^M_{(c)} A_M^{(c)} \equiv dx^\mu {\cal B}_\mu^{(c)} +
d\bar\theta {\cal F}^{(c)}
\end{array}\eqno(3.3)
$$ where $Z^M_{(c)} = (x^\mu, \bar\theta)$ is the chiral
superspace variable, $\partial^{(c)}_M$ is the chiral partial
derivative and $A_M^{(c)} = ({\cal B}_\mu^{(c)}, {\cal F}^{(c)})$
is the chiral supermultiplet. The r.h.s. of (3.2) leads to the
definition of the ordinary 2-form $F^{(2)}$ (as well as the field
strength tensor $F_{\mu\nu}$) as: $$
\begin{array}{lcl}
\phi^* (x) \; D\; D\; \phi (x) = i e \phi^* (x)\; (F^{(2)})\; \phi
(x) \equiv \frac{1}{2!} (dx^\mu \wedge dx^\nu)\; \phi^* (x)
(F_{\mu\nu})\; \phi (x).
\end{array}\eqno(3.4)
$$ It is straightforward to check that the above quantity is a
$U(1)$ gauge (i.e. BRST) invariant quantity because $\tilde s_b
\phi = - i e C \phi, \tilde s_b F_{\mu\nu} = 0, \tilde s_b \phi^*
= + i e \phi^* C \Rightarrow \tilde s_b (\phi^* F_{\mu\nu} \phi) =
0$.

The l.h.s. of the gauge invariant restriction (3.2) would yield
the coefficients of the 2-form differentials $(dx^\mu \wedge
dx^\nu), (dx^\mu \wedge d\bar\theta), (d\bar\theta \wedge
d\bar\theta)$. It is evident, on the other hand, that the r.h.s.
of the restriction yields only the coefficients of $(dx^\mu \wedge
dx^\nu)$ (cf. (3.4)). The expanded version of the l.h.s. of the
restriction in (3.2) is $$
\begin{array}{lcl}
&& (dx^\mu \wedge dx^\nu)\; \Phi^*_{(c)} (x, \bar\theta)\; \bigl [
\;(\partial_\mu + i e {\cal B}_\mu^{(c)})\;(\partial_\nu + i e
{\cal B}^{(c)}_\nu)\; \bigr ]\;\Phi_{(c)} (x, \bar\theta) - (d
\bar\theta \wedge d\bar\theta)\nonumber\\ && \Phi^*_{(c)} (x,
\bar\theta)\; \bigl [ \;(\partial_{\bar\theta} + i e {\cal
F}^{(c)})\; (\partial_{\bar\theta} + i e {\cal F}^{(c)})\; \bigr
]\; \Phi_{(c)} (x, \bar\theta) +  (dx^\mu \wedge
d\bar\theta)\nonumber\\ && \Phi^*_{(c)}  (x, \bar\theta) \;\bigl [
(\partial_\mu + i e {\cal B}_\mu^{(c)})\; (\partial_{\bar\theta} +
i e {\cal F}^{(c)}) - (\partial_{\bar\theta} + i e {\cal F}^{(c)})
(\partial_\mu + i e {\cal B}^{(c)}_\mu) \bigr ]\; \Phi_{(c)} (x,
\bar\theta).
\end{array}\eqno(3.5) $$
It is obvious that the coefficients of the 2-form differentials
$(d \bar\theta \wedge d \bar\theta), (dx^\mu \wedge d\bar\theta)$
would be set equal to zero to maintain the sanctity of (3.2). Such
an operation on the coefficient of the former, leads to $$
\begin{array}{lcl} -\; i\; e
\;\Phi^*_{(c)} (x, \bar\theta)\; (\partial_{\bar\theta} {\cal
F}^{(c)})\;\Phi_{(c)} (x, \bar\theta) = 0.
\end{array}\eqno(3.6)
$$ For $e \neq 0, \Phi_{(c)} \neq 0, \Phi^*_{(c)} \neq 0$, we
obtain $\partial_{\bar\theta} {\cal F}^{(c)} = 0$ which implies
$B_1 (x) = 0$ in the expansion of ${\cal F}^{(c)} (x, \bar\theta)$
in (3.1). This shows that the reduced form (i.e. ${\cal F}^{(c)}
(x, \bar\theta)  \to {\cal F}^{(c)}_{(r)} (x, \bar\theta)$)  of
the expansion for the fermionic chiral superfield ${\cal F}^{(c)}
(x, \bar\theta)$ is: ${\cal F}^{(c)}_{(r)} (x, \bar\theta) = C
(x)$. This leads to primarily a pair of consequences. First, to
maintain the sanctity of the restriction (3.2), the chiral
superfield  ${\cal F}^{(c)} (x, \bar\theta)$ becomes a local
ordinary 4D field $C (x)$. Second, it implies that the on-shell
nilpotent BRST transformations for the ghost field $C(x)$ is zero
(i.e. ${\cal F}^{(c)}_{(r)} (x, \bar\theta) = C (x) + \bar\theta\;
(\tilde s_b C (x))$). Setting the coefficient of $(dx^\mu \wedge d
\bar\theta)$ equal to zero, ultimately, implies the following
relationship between the chiral superfields: $$
\begin{array}{lcl}
\partial_\mu {\cal F}^{(c)}_{(r)} = \partial_{\bar\theta} {\cal
B}^{(c)}_\mu  \; \;\Rightarrow\; \; R_\mu (x) = \partial_\mu C (x)
\end{array}\eqno(3.7)
$$ when $e \neq 0, \Phi_{(c)} \neq 0, \Phi^*_{(c)} \neq 0$. Thus,
the reduced form (i.e. ${\cal B}^{(c)}_\mu (x, \bar\theta) \to
{\cal B}^{(c)}_{\mu (r)} (x, \bar\theta)$) of the bosonic
superfield, after the application of the restriction (3.2),
becomes $$
\begin{array}{lcl}
{\cal B}^{(c)}_{\mu (r)} (x, \bar\theta) = A_\mu (x) \;+
\;\bar\theta\;\partial_\mu C (x) \equiv A_\mu (x)\; +\;
\bar\theta\; (\tilde s_b A_\mu (x)).
\end{array}\eqno(3.8)
$$ The above equation demonstrates the derivation of the on-shell
nilpotent BRST symmetry transformation for the gauge field $A_\mu$
within the framework of superfield formalism.

Finally, we equate the coefficient of $(dx^\mu \wedge dx^\nu)$
from the l.h.s. and r.h.s. of the gauge (i.e. BRST) invariant
restriction (3.2). The precise form of this equality is$$
\begin{array}{lcl}
&& \frac{1}{2}\; i e \;(dx^\mu \wedge dx^\nu)\; \Phi^*_{(c)} (x,
\bar\theta) \; \bigl (\partial_\mu {\cal B}^{(c)}_{\nu (r)} -
\partial_\nu {\cal B}^{(c)}_{\mu (r)} \bigr ) \; \Phi_{(c)} (x,
\bar\theta) \nonumber\\ && = \frac{1}{2}\; i e\; (dx^\mu \wedge
dx^\nu)\; \phi^* (x) \; \bigl (\partial_\mu A_\nu -
\partial_\nu A_\mu \bigr )\; \phi (x).
\end{array}\eqno(3.9)
$$ Using (3.8), it is straightforward to note that $\partial_\mu
{\cal B}^{(c)}_{\nu(r)} - \partial_\nu {\cal B}^{(c)}_{\mu(r)} =
\partial_\mu A_\nu - \partial_\nu A_\mu$. Ultimately, the above
equality in (3.9) reduces to the following form\footnote{It will
be noted that the condition in (3.10) is a completely new
relationship which can never originate from the horizontality
condition alone. In fact, the horizontality condition, present in
the usual superfield formalism [1-7], does not shed any light on
the derivation of the (anti-)BRST symmetry transformations for the
matter fields of a given interacting gauge theory, as pointed out
earlier.} $$
\begin{array}{lcl}
\Phi^*_{(c)} (x, \bar\theta) \; \Phi_{(c)} (x, \bar\theta) =
\phi^* (x) \; \phi (x).
\end{array}\eqno(3.10)
$$ The above simplicity occurs because of the Abelian nature of
the gauge theory under consideration. The same does not hold good
for the non-Abelian interacting gauge theory where the gauge
fields are group valued and, therefore, noncommutative in nature
(see, e.g., [19] for details). The substitution of the expansions
in (3.1) for the chiral superfields on the l.h.s. of (3.10) leads
to the following condition $$
\begin{array}{lcl}
\phi^* (x) \; f_1 (x) + f_2^* (x)\; \phi (x) = 0.
\end{array}\eqno(3.11)
$$ One of the simplest solutions to the above condition is the
case where $f_1 (x)$ is proportional to the basic field $\phi (x)$
and $f_2^* (x)$ is that of the 4D field $\phi^* (x)$. However, it
should be noted that the secondary fields $f_1 (x)$ and $f_2^*
(x)$ are fermionic in nature whereas the complex scaler fields
$\phi (x)$ and $\phi^* (x)$ are bosonic. For the precise value of
the equality, one of the interesting choices (that makes sense)
is: $$
\begin{array}{lcl}
f_1 (x) = - e C (x)  \phi (x)\; \qquad \;f_2^* (x) = + e \; \phi^*
(x) C (x)
\end{array}\eqno(3.12)
$$ where field $C (x)$ is the fermionic ghost field of the theory.
This field has been brought in to make the above choice fermionic
in nature for $f_1 (x)$ and $f_2^* (x)$. The above choices, in
some sense, are unique because the presence of the fermionic ghost
field $C(x)$ is the only appropriate possibility in (3.12). The
substitution of the above values into the super expansion (3.1)
leads to the derivation of $\tilde s_b$ (cf. (2.5)) for the matter
fields as given below $$
\begin{array}{lcl}
&&\Phi^{(r)}_{(c)} (x, \bar\theta) = \phi (x) + \bar\theta\; (-i e
C (x) \phi (x)) \equiv \phi (x) + \bar\theta \; (\tilde s_b \phi
(x)) \nonumber\\ && \Phi^{*(r)}_{(c)} (x, \bar\theta) =  \phi^*
(x) + \bar\theta \;(+ i e \phi^* (x) C (x)) \equiv \phi^* (x) +
\bar\theta\; (\tilde s_b \phi^* (x)).
\end{array}\eqno(3.13)
$$ It should be emphasized that, so far, we have not been able to
determine the exact value of the secondary field $B_2 (x)$ in
terms of the basic fields of the theory by exploiting the gauge
(i.e. BRST) invariant  restriction (3.2). At this stage, the
equation of motion $B = - (\partial \cdot A)$ derived from (2.1)
(or (2.3)) comes to our rescue if we identify \footnote{We lay
stress, at this point of our argument, that we shall remain
consistent with this identification [i.e. $ B_2 (x) = B (x) \equiv
- (\partial \cdot A)$] throughout the body of our present text.}
the secondary field $B_2 (x)$ with the Nakanishi-Lautrup auxiliary
field $B(x)$. With this input, we obtain the on-shell nilpotent
symmetry transformations for all basic fields of the theory as the
expansion (3.1) can be re-expressed, in terms of $\tilde s_b$ (cf.
(2.5)), as $$
\begin{array}{lcl}
{\cal B}^{(c)}_{\mu(r)} (x, \bar\theta) &=& A_\mu (x) +
\bar\theta\; (\tilde s_b A_\mu (x))\; \qquad \;{\cal
F}^{(c)}_{(r)} (x, \bar\theta) = C (x) + \; \bar\theta\; (\tilde
s_b C (x)) \nonumber\\ \bar {\cal F}^{(c)}_{(r)} (x, \bar\theta) &
= & C (x) +  \; \bar\theta\; (\tilde s_b \bar C (x)) \;\qquad\;\;
\Phi^{(r)}_{(c)} (x, \bar\theta)  =  \phi (x) +  \; \bar\theta\;
(\tilde s_b \phi (x)) \nonumber\\ \Phi^{*(r)}_{(c)} (x,
\bar\theta) &=& \phi^* (x) +  \; \bar\theta\; (\tilde s_b\phi^*
(x)).
\end{array}\eqno(3.14)
$$ The above equation provides the geometrical interpretation for
the on-shell nilpotent BRST transformation $\tilde s_b$ (and the
corresponding generator $\tilde Q_b$) as the translational
generator along the Grassmannian direction $\bar\theta$ of the
chiral super sub-manifold. In other words, the translation of the
(4, 1)-dimensional chiral  superfields along the
$\bar\theta$-direction of the chiral super sub-manifold results in
the internal on-shell nilpotent BRST symmetry transformations
$\tilde s_b$ for the corresponding basic 4D local fields of the
Lagrangian density (2.6).\\

\noindent {\bf 3.2 On-shell nilpotent anti-BRST symmetries:
anti-chiral superfields}\\

\noindent We invoke here the anti-chiral (i.e. $\bar \theta = 0$)
superfields ${\cal B}_\mu^{(ac)}, {\cal F}^{(ac)}, \bar {\cal
F}^{(ac)}, \Phi_{(ac)}, \Phi^*_{(ac)}$, corresponding to the basic
fields $A_\mu, C, \bar C, \phi, \phi^*$ of the 4D Lagrangian
density (2.6), for the derivation of the anti-BRST symmetry
transformations of (2.5). We expand the above superfields along
the $\theta$-direction of the (4, 1)-dimensional anti-chiral super
sub-manifold. These expansions are $$
\begin{array}{lcl}
{\cal B}^{(ac)}_\mu (x, \theta) &=& A_\mu (x) + \theta\; \bar
R_\mu (x)\; \qquad \;{\cal F}^{(ac)} (x, \theta) = C (x) + i \;
\theta\; \bar B_1 (x) \nonumber\\ \bar {\cal F}^{(ac)} (x, \theta)
& = & C (x) + i \; \theta\; \bar B_2 (x)\; \qquad \;\Phi_{(ac)}
(x, \theta)  =  \phi (x) + i \; \theta\; \bar f_1 (x) \nonumber\\
\Phi^*_{(ac)} (x, \theta) &=& \phi^* (x) + i \; \theta\; \bar
f^*_2 (x)
\end{array}\eqno(3.15)
$$ where the basic fields the Lagrangian density (2.6) are
obtained in the limit $\theta \to 0$. In the above expansion, the
fields  $\bar R_\mu, \bar B_1, \bar B_2, \bar f_1, \bar f_2^*$ are
the secondary fields which would be determined in terms of the
basic fields of (2.6) by the imposition of the following gauge
(i.e. (anti-)BRST) invariant restriction \footnote{ It is
interesting to note that the combination $\Phi_{(ac)} (x, \theta)
\tilde {\bar {\cal D}}_{(ac)}\; \tilde {\bar {\cal D}}_{(ac)}
\Phi^*_{(ac)} (x, \theta) = \phi (x) \bar D \bar D \phi^* (x)$ is
also a gauge (i.e. (anti-)BRST) invariant condition that could be
imposed on the anti-chiral superfields of the (4, 1)-dimensional
super sub-manifold. This restriction also leads to the derivation
of the on-shell nilpotent anti-BRST symmetry transformations for
all the fields of the theory. The computational steps are similar
to those connected with the present condition in (3.16).} on the
anti-chiral superfields defined on the (4, 1)-dimensional super
sub-manifold of the general (4, 2)-dimensional supermanifold;
namely; $$
\begin{array}{lcl}
\Phi^*_{(ac)} (x, \theta)\; \tilde {\cal D}_{(ac)}\; \tilde {\cal
D}_{(ac)} \Phi_{(ac)} (x, \theta) = \phi^* (x)\; D\;D\; \phi (x)
\end{array}\eqno(3.16)
$$ where the anti-chiral covariant derivative $\tilde {\cal
D}_{(ac)}$, on the (4, 1)-dimensional anti-chiral super
sub-manifold, is defined as $$
\begin{array}{lcl}
\tilde {\cal D}_{(ac)} = \tilde d_{(ac)} + i e \tilde
A^{(1)}_{(ac)} \equiv dx^\mu (\partial_\mu + i e {\cal
B}^{(ac)}_\mu) + d \theta (\partial_\theta + i e \bar {\cal
F}^{(ac)}).
\end{array}\eqno(3.17)
$$ Here $\tilde d^{(ac)} = dx^\mu \partial_\mu + d \theta
\partial_\theta$ is the anti-chiral version of the super exterior
derivative $\tilde d = dx^\mu \partial_\mu + d \theta
\partial_\theta + d \bar\theta \partial_{\bar\theta}$  and
$\tilde A^{(1)}_{(ac)} = dx^\mu {\cal B}^{(ac)}_\mu + d \theta
\bar {\cal F}^{(ac)}$ is the anti-chiral limit of the super 1-form
connection $\tilde A^{(1)} = dx^\mu {\cal B}_\mu (x, \theta,
\bar\theta) + d \theta \bar {\cal F} (x, \theta, \bar\theta) + d
\bar\theta {\cal F} (x, \theta, \bar\theta)$ that would be
exploited in the next section (cf. Sec. 4) for the derivation of
the off-shell nilpotent (anti-)BRST symmetry transformations for
all the fields of the theory. It is straightforward to note that
the r.h.s. of (3.16) defines the 2-form $F^{(2)}$ as: $ i e \phi^*
(x) F^{(2)} \phi (x) \equiv \frac{1}{2} i e (dx^\mu \wedge dx^\nu)
\phi^* (x) F_{\mu\nu} \phi (x) $.

Let us focus on the explicit form of the l.h.s. of (3.16). In
terms of the 2-form differentials $(dx^\mu \wedge dx^\nu), (dx^\mu
\wedge d\theta)$ and $(d \theta \wedge d\theta)$, the l.h.s. can
be written in its most lucid form as $$
\begin{array}{lcl}
&& (dx^\mu \wedge dx^\nu)\; \Phi^*_{(ac)} (x, \theta)\; \bigl [
\;(\partial_\mu + i e {\cal B}_\mu^{(ac)})\;(\partial_\nu + i e
{\cal B}^{(ac)}_\nu)\; \bigr ]\;\Phi_{(ac)} (x, \theta) - (d
\theta \wedge d\theta)\nonumber\\ && \Phi^*_{(ac)} (x, \theta)\;
\bigl [ \;(\partial_{\theta} + i e \bar {\cal F}^{(ac)})\;
(\partial_{\theta} + i e \bar {\cal F}^{(ac)})\; \bigr ]\;
\Phi_{(ac)} (x, \theta) +  (dx^\mu \wedge d\theta)\; \Phi^*_{(ac)}
(x, \theta) \nonumber\\ && \bigl [ (\partial_\mu + i e {\cal
B}_\mu^{(ac)})\; (\partial_{\theta} + i e \bar {\cal F}^{(ac)}) -
(\partial_{\theta} + i e \bar {\cal F}^{(ac)}) (\partial_\mu + i e
{\cal B}^{(ac)}_\mu) \bigr ]\; \Phi_{(ac)} (x, \theta).
\end{array}\eqno(3.18) $$
It is evident that the coefficient of $(d \theta \wedge d\theta)$
of the above equation would be set equal to zero because there is
no such term on the r.h.s. of (3.16). The simplified version of
the consequence of this statement can be expressed as $$
\begin{array}{lcl}
- i\; e\; (d \theta \wedge d\theta) \;\Phi^*_{(ac)} (x, \theta)\;
\bigl (\partial_\theta \bar {\cal F}^{(ac)} \bigr )\; \Phi_{(ac)}
(x, \theta) = 0.
\end{array}\eqno(3.19)
$$ For $ e \neq 0, \Phi \neq 0, \Phi^* \neq 0$, we obtain the
solution $\partial_\theta \bar {\cal F}^{(ac)} = 0$. This leads to
$\bar B_2 = 0$ in the expansion of $\bar {\cal F}^{(ac)} (x,
\theta)$. Thus, the reduced form (i.e. $\bar {\cal F}^{(ac)} \to
\bar {\cal F}^{(ac)}_{(r)}$) of this superfield can be
re-expressed, in terms of the on-shell nilpotent operator $\tilde
s_{ab}$,  as $$
\begin{array}{lcl}
\bar {\cal F}^{(ac)}_{(r)} (x, \theta) = \bar C (x) + 0 \equiv
\bar C (x) + \theta \; (\tilde s_{ab} \bar C(x)).
\end{array}\eqno(3.20)
$$ The above equation demonstrates the explicit derivation of the
anti-BRST symmetry transformations for the anti-ghost field as
$\tilde s_{ab} \bar C (x) = 0$.

We now collect the coefficient of the 2-form differential $(dx^\mu
\wedge d\theta)$ from the equation (3.18). In its simple form, it
looks as follows $$
\begin{array}{lcl}
i e (dx^\mu \wedge d\theta)\; \Phi^*_{(ac)} (x, \theta) \bigl
(\partial_\mu \bar {\cal F}^{(ac)}_{(r)} - \partial_\theta {\cal
B}^{(ac)}_\mu \bigr )\; \Phi_{(ac)} (x, \theta) = 0.
\end{array}\eqno(3.21)
$$ It is obvious that for $e \neq 0, \Phi_{(ac)} \neq 0,
\Phi^*_{(ac)} \neq 0$, we obtain the relationship $\partial_\mu
\bar {\cal F}^{(ac)}_{(r)} = \partial_\theta {\cal B}^{(ac)}_\mu$
which implies that $\bar R_\mu (x) = \partial_\mu \bar C (x)$.
Thus, the reduced form (i.e. ${\cal B}_\mu^{(ac)} (x, \theta) \to
{\cal B}^{(ac)}_{\mu(r)} (x, \theta)$) of the bosonic superfield
(corresponding to the gauge field $A_\mu$)  is $$
\begin{array}{lcl}
{\cal B}^{(ac)}_{\mu(r)} (x, \theta) = A_\mu (x) + \theta\;
\partial_\mu \bar C (x) \equiv A_\mu (x) + \theta\; (\tilde s_{ab}
A_\mu (x)).
\end{array}\eqno(3.22)
$$ The above equation establishes the exact derivation of the
nilpotent anti-BRST symmetry transformation for the gauge field
$A_\mu$ in the framework of the present superfield formalism.
Collecting the coefficient of $(dx^\mu \wedge dx^\nu)$ from both
the sides of the restriction (3.16), we obtain the following
relationship $$
\begin{array}{lcl}
&& \frac{1}{2}\; i e \;(dx^\mu \wedge dx^\nu)\; \Phi^*_{(ac)} (x,
\theta) \; \bigl (\partial_\mu {\cal B}^{(ac)}_{\nu (r)} -
\partial_\nu {\cal B}^{(ac)}_{\mu (r)} \bigr ) \; \Phi_{(ac)} (x,
\theta) \nonumber\\ && = \frac{1}{2}\; i e\; (dx^\mu \wedge
dx^\nu)\; \phi^* (x) \; \bigl (\partial_\mu A_\nu -
\partial_\nu A_\mu \bigr )\; \phi (x).
\end{array}\eqno(3.23)
$$ Taking the help of (3.22), it is evident that $\partial_\mu
{\cal B}^{(ac)}_{\nu(r)} - \partial_\nu {\cal B}^{(ac)}_{\mu(r)} =
\partial_\mu A_\nu - \partial_\nu A_\mu$. Thus, the simplest form
of the condition in (3.23), that emerges after a bit of simple
algebra, is: $\Phi^*_{(ac)} (x, \theta) \Phi_{(ac)} (x, \theta) =
\phi^* (x) \phi (x)$. It will be noted that this new relation is a
gauge (i.e. BRST) invariant relation and it cannot emerge from the
usual horizontality condition. Inserting the expansion of (3.15)
for the anti-chiral matter fields, we obtain the following
relationship among the basic fields $\phi, \phi^*$ and the
secondary fermionic fields $\bar f_1$ and $\bar f_2^*$, namely; $$
\begin{array}{lcl}
\phi^* (x)\; \bar f_1 (x) + \bar f_2^* (x) \phi (x) = 0.
\end{array}\eqno(3.24)
$$ A close look at the above condition provides us the clue to
choose the secondary fermionic fields $\bar f_1$ proportional to
$\phi$ and $\bar f_2^*$ proportional to $\phi^*$. For the exact
equality, we bring in the anti-ghost field $\bar C (x)$ of the
theory which allows us to choose the following $$
\begin{array}{lcl}
\bar f_1 (x) = - e\; \bar C (x) \; \phi (x) \;\; \qquad \;\;\bar
f_2^* (x) = + e\; \phi^* (x) \; \bar C (x).
\end{array}\eqno(3.25)
$$ The insertion of the above values into the expansion (3.15)
leads to the exact derivation of the anti-BRST symmetry
transformations for the matter fields in the sense that the
reduced form of the matter fields become: $\Phi_{(ac)}^{(r)} (x,
\theta) = \phi (x) + \theta\; (\tilde s_{ab} \phi (x)),
\Phi_{(ac)}^{*(r)} (x, \theta) = \phi^* (x) + \theta\; (\tilde
s_{ab} \phi^* (x))$ in terms of the on-shell nilpotent anti-BRST
symmetry transformations.

So far, we have not been able to determine the secondary field
$\bar B_1 (x)$ of the expansion of ${\cal F}^{(ac)} (x, \theta)$
(cf (3.15) in terms of the basic fields of the Lagrangian density
(2.6). In fact, $\bar B_1 (x)$ can be identified with the
Nakanishi-Lautrup auxiliary field $B (x)$ with a minus sign (i.e.
$\bar B_1 (x) = - B (x)$). The reason behind this choice with a
minus sign will become clear in the next section. We shall be
consistent, however, with this specific choice throughout the body
of our present text. At this stage, once again, the equation of
motion $B (x) = - (\partial \cdot A) (x)$ comes to our help. Thus,
the exact expression for the secondary field $\bar B_1 (x)$
becomes $\bar B_1 (x) = + (\partial \cdot A) (x)$. Insertion of
this value in the expansion of ${\cal F}^{(ac)} (x, \theta)$ (cf.
(3.15)) reduces this superfield to the form ${\cal F}^{(ac)}_{(r)}
(x, \theta) = C (x) + i \theta (\partial \cdot A) \equiv C (x) +
\theta\; (\tilde s_{ab} C(x))$. Ultimately, all the superfields in
their reduced form (with the inputs from the on-shell nilpotent
anti-BRST transformations (2.5)) can be re-expressed as$$
\begin{array}{lcl}
{\cal B}^{(ac)}_{\mu(r)} (x, \theta) &=& A_\mu (x) + \;\theta\;
(\tilde s_{ab} A_\mu (x)) \qquad {\cal F}^{(ac)}_{(r)} (x, \theta)
= C (x) + \;\theta\; (\tilde s_{ab} C (x)) \nonumber\\ \bar {\cal
F}^{(ac)}_{(r)} (x, \theta) & = & C (x) + \;\theta\; (\tilde
s_{ab} \bar C (x))\; \qquad \;\Phi^{(r)}_{(ac)} (x, \theta)  =
\phi (x) + \; \theta\; (\tilde s_{ab} \phi (x)) \nonumber\\
\Phi^{*(r)}_{(ac)} (x, \theta) &=& \phi^* (x) + \; \theta\;
(\tilde s_{ab}\phi^* (x)).
\end{array}\eqno(3.26)
$$ The above set of expansions provides the geometrical origin and
interpretation for the on-shell nilpotent anti-BRST symmetry
transformation (and corresponding generator $\tilde Q_{ab}$) as
the translational generator along $\theta$-direction of the
anti-chiral super sub-manifold.\\

\noindent {\bf 4 Off-shell nilpotent (anti-)BRST symmetries:
superfield formalism}\\

\noindent To obtain the off-shell nilpotent symmetry
transformations (2.4) for all the fields of the theory in
superfield formalism, we define the 4D ordinary interacting $U(1)$
gauge theory with complex scalar fields on a $(4, 2)$-dimensional
supermanifold parametrized by the general superspace coordinate
$Z^M = (x^\mu, \theta, \bar \theta)$ where $x^\mu (\mu = 0, 1, 2,
3)$ are the four even spacetime coordinates and $\theta, \bar
\theta$  are a pair of odd elements of a Grassmann algebra. On
this supermanifold, one can define a  set of superfields
corresponding to the basic fields of the theory that are present
in the Lagrangian density (2.6). The above superfields can be
expanded in terms of these basic fields $A_\mu, C, \bar C, \phi,
\phi^*$ and some secondary fields (along the Grassmannian
directions of the (4, 2)-dimensional supermanifold) as [3,13,14]
$$
\begin{array}{lcl}
{\cal B}_{\mu} (x, \theta, \bar \theta) &=& A_{\mu} (x) + \theta\;
\bar R_{\mu} (x) + \bar \theta\; R_{\mu} (x) + i \;\theta \;\bar
\theta \;S_{\mu} (x) \nonumber\\ {\cal F} (x, \theta, \bar \theta)
&=& C (x) + i\; \theta \bar B_1 (x) + i \;\bar \theta\; B_1 (x) +
i\; \theta\; \bar \theta \;s (x) \nonumber\\ \bar {\cal F}  (x,
\theta, \bar \theta) &=& \bar C (x) + i \;\theta\;\bar B_2 (x) +
i\; \bar \theta \;B_2 (x) + i \;\theta \;\bar \theta \;\bar s (x)
\nonumber\\ \Phi (x, \theta, \bar\theta) &=& \phi (x) + i\;
\theta\; \bar f_1 (x) + i\; \bar\theta\; f_1 (x) + i\; \theta\;
\bar\theta\; b(x) \nonumber\\ \Phi^* (x, \theta, \bar\theta) & = &
\phi^* (x) + i \;\theta\; \bar f_2^* (x) + i\;\bar\theta\; f^*_2
(x) + i \;\theta \;\bar\theta\;  b^*(x).
\end{array} \eqno(4.1)
$$ It is straightforward to note, in the above super expansion,
that the local fields $ R_{\mu} (x)$, $\bar R_{\mu} (x)$, $C (x)$,
$\bar C (x)$, $s (x)$, $\bar s (x)$,$ f_1 (x)$, $\bar f_1 (x)$,
$f_2^* (x)$, $\bar f_2^* (x) $ are fermionic (anticommuting) and
$A_{\mu} (x), S_{\mu} (x), B_1 (x), \bar B_1 (x), B_2 (x), \bar
B_2 (x)$ are bosonic (commuting) in nature. In the above
expansion, the bosonic-
 and fermionic degrees of freedom match and, in the limit:
$\theta, \bar\theta \rightarrow 0$, we get back our basic fields
$A_\mu, C, \bar C, \phi, \phi^*$ of the Lagrangian density (2.6)

To obtain the exact expressions for the secondary fields in terms
of the basic fields (and auxiliary fields) of the theory, we
invoke the following gauge (i.e. (anti-)BRST) invariant
restriction on the suitable superfields of the general (4,
2)-dimensional supermanifold, namely;

$$ \begin{array}{lcl} \Phi^* (x, \theta, \bar\theta) \; \tilde
{\cal D}\; \tilde {\cal D}\; \Phi (x, \theta, \bar\theta) = \phi^*
(x) \; D \; D \;\phi (x)
\end{array}\eqno(4.2)
$$ where the super covariant derivative $\tilde {\cal D}$ is
defined, in terms of the super exterior derivative $\tilde d =
dx^\mu \partial_\mu + d \theta \partial_\theta + d \bar\theta
\partial_{\bar\theta}$ and the 1-form super connection
$\tilde A^{(1)} = d x^\mu {\cal B}_\mu + d \theta \bar {\cal F} +
d \bar \theta {\cal F}$, as $$
\begin{array}{lcl}
\tilde {\cal D} = \tilde d + i e \tilde A^{(1)} \equiv dx^\mu\;
(\partial_\mu + i e {\cal B}_\mu) + d \theta \;(\partial_\theta +
i e \bar {\cal F}) + d \bar\theta\; (\partial_{\bar\theta} + i e
{\cal F}).
\end{array}\eqno(4.3)
$$ It will be noted that, in the previous section, we have taken
the limiting cases (i.e $\theta \to 0$ and $\bar\theta \to 0$) of
the above definition for the chiral  and anti-chiral super
sub-manifolds. It is evident that the r.h.s. of (4.2) (i.e. $i \;e
\;\phi^* (x)\; F^{(2)}\; \phi (x)$)  is a gauge (i.e. (anti-)BRST)
invariant quantity on the supermanifold because $s_{(a)b}\;
(\phi^* F^{(2)} \phi) = 0$ (as is clear from the off-shell
nilpotent (anti-)BRST transformations $s_{(a)b}$ quoted in
equation (2.4)).

For our computations, it is important to express the l.h.s. of the
gauge invariant restriction (4.2) in terms of the explicit 2-form
differentials $(dx^\mu \wedge dx^\nu), (d \theta \wedge d\theta),
(d \bar\theta \wedge d\bar \theta), (d\theta \wedge d\bar\theta),
(dx^\mu \wedge d\theta)$ and $(dx^\mu \wedge d\bar\theta)$. This
is required so that we can compare the r.h.s. with the l.h.s. of
(4.2). All the terms, corresponding to these differentials
together with their coefficients from the l.h.s. of the
restriction (4.2), are explicitly written as follows $$
\begin{array}{lcl}
&& (dx^\mu \wedge dx^\nu)\; \Phi^* \;\bigl (\partial_\mu + i e
{\cal B}_\mu) (\partial_\nu + i e {\cal B}_\nu \bigr )\; \Phi
\nonumber\\ && - (d\theta \wedge d\theta)\; \Phi^* \;\bigl
(\partial_\theta + i e \bar {\cal F} \bigr )\;\bigl
(\partial_\theta + i e \bar {\cal F} \bigr ) \; \Phi \nonumber\\
&& - (d\bar\theta \wedge d\bar\theta)\; \Phi^* \;\bigl
(\partial_{\bar\theta} + i e  {\cal F} \bigr )\;\bigl
(\partial_{\bar\theta} + i e  {\cal F} \bigr ) \; \Phi \nonumber\\
&& - (d\theta \wedge d\bar\theta)\; \Phi^* \;\bigl [
(\partial_{\bar\theta} + i e  {\cal F}) (\partial_\theta + i e
\bar {\cal F}) +  (\partial_\theta + i e \bar {\cal F})
(\partial_{\bar\theta} + i e {\cal F}) \bigr ] \; \Phi \nonumber\\
&& + (dx^\mu \wedge d\theta)\; \Phi^*\; \bigl [ \;(\partial_\mu +
i e {\cal B}_\mu) (\partial_\theta + i e \bar {\cal F}) -
(\partial_\theta + i e \bar {\cal F}) (\partial_\mu + i e {\cal
B}_\mu) \bigr ]\; \Phi \nonumber\\ && + (dx^\mu \wedge
d\bar\theta)\; \Phi^*\; \bigl [ \;(\partial_\mu + i e {\cal
B}_\mu) (\partial_{\bar\theta} + i e  {\cal F}) -
(\partial_{\bar\theta} + i e  {\cal F}) (\partial_\mu + i e {\cal
B}_\mu) \bigr ]\; \Phi.
\end{array}\eqno(4.4)
$$ For algebraic convenience, first of all, it is useful to set
equal to zero the coefficients of the differentials $(d\theta
\wedge d\theta)$, $(d \bar\theta \wedge d\bar\theta)$  and $(d
\theta \wedge d\bar\theta)$ as these are not present on the r.h.s.
of the restriction (4.2). The outcome of the above algebraic
conditions are $$
\begin{array}{lcl}
&&- i\; e\; (d\theta \wedge d\theta)\; \Phi^* \;(\partial_\theta
\bar {\cal F})\; \Phi = 0\; \qquad\; - i\; e \;(d\bar\theta \wedge
d\bar\theta)\; \Phi^* \;(\partial_{\bar \theta}  {\cal F})\; \Phi
= 0 \nonumber\\ && - \;i\; e\; (d \theta \wedge d\bar\theta)\;
\Phi^* \; \bigl (\partial_\theta {\cal F} + \partial_{\bar\theta}
\bar {\cal F} \bigr )\; \Phi = 0.
\end{array}\eqno(4.5)
$$ For $e \neq 0, \Phi \neq 0, \Phi^* \neq 0$, we obtain the
following solutions $$
\begin{array}{lcl}
&&\partial_\theta \bar {\cal F} = 0 \;\Rightarrow\;  \bar B_2 (x)
= 0\; \qquad \;\bar s (x) = 0 \nonumber\\ && \partial_{\bar\theta}
{\cal F} = 0 \;\Rightarrow  \; B_1 (x) = 0\; \qquad \;s (x) =
0,\nonumber\\ &&
\partial_\theta {\cal F} + \partial_{\bar\theta} \bar {\cal F} = 0
\;\;\Rightarrow \;\;B_2 (x) + \bar B_1 (x) = 0.
\end{array}\eqno(4.6)
$$ The insertions of these values into the super expansion of
${\cal F}$ and $\bar {\cal F}$ (along with our earlier
identifications: $ B_2 (x) = B (x), \bar B_1 (x) = - B(x)$), imply
the following reduced forms (i.e. ${\cal F} (x, \theta,
\bar\theta) \to {\cal F}_{(r)} (x, \theta), \bar {\cal F} (x,
\theta, \bar\theta) \to \bar {\cal F}_{(r)} (x, \bar\theta)$) of
the superfield expansions $$
\begin{array}{lcl}
&& {\cal F}_{(r)} (x, \theta) = C (x) - i\; \theta\; B (x) \equiv
C (x) + \theta\; (s_{ab} C (x)) \nonumber\\ && \bar {\cal F}_{(r)}
(x, \bar\theta) = \bar C (x) + i \; \bar\theta\; B (x) \equiv \bar
C (x) + \bar\theta\; (s_b \bar C (x)).
\end{array}\eqno(4.7)
$$ The above equation imply (i) the derivation of the off-shell
nilpotent  (anti-)BRST symmetry transformations for the ghost and
anti-ghost fields of the theory under consideration, (ii) the
characteristic features of the  superfields $\bar {\cal F}$ and
${\cal F}$ as the chiral and anti-chiral in nature after the
application of the restriction (4.2), and (iii) the choices made
in the previous section are correct because it can be seen that if
$B_2 (x) = B (x)$, the relation $B_2 (x) + \bar B_1 (x) = 0$
implies that $\bar B_1 (x) = - B(x)$. These results, which have
been discussed above, are exactly same as that derived due to the
application of the horizontality condition of the usual superfield
formalism on the (4, 2)-dimensional supermanifold (see, e.g,
[3,8,9] for details).

We now collect the coefficients of the 2-form differentials
$(dx^\mu \wedge d \theta)$ and $(dx^\mu \wedge d\bar\theta)$.
These are, naturally, to be set equal to zero. The consequences
are listed below $$
\begin{array}{lcl}
&& + \;i\; e\; (dx^\mu \wedge d\theta)\; \Phi^*\; \bigl
(\partial_\mu \bar {\cal F}_{(r)}  - \partial_\theta {\cal B}_\mu
\bigr )\; \Phi = 0 \nonumber\\ && +\; i\; e\; (dx^\mu \wedge d
\bar \theta)\; \Phi^*\; \bigl (\partial_\mu  {\cal F}_{(r)}  -
\partial_{\bar \theta} {\cal B}_\mu \bigr )\; \Phi = 0.
\end{array}\eqno(4.8)
$$ It will be noted here that the reduced values (4.7) of the
superfields ${\cal F}$ and $\bar {\cal F}$ have been taken into
account for the above computations. For $e \neq 0, \Phi \neq 0,
\Phi^* \neq 0$, we obtain the following explicit and precise
solutions to the above restrictions $$
\begin{array}{lcl}
R_\mu (x) = \partial_\mu C (x)\; \qquad \;\bar R_\mu (x) =
\partial_\mu \bar C (x)\; \qquad \;S_\mu (x) = \partial_\mu B (x).
\end{array}\eqno(4.9) $$ The insertions
of the above values into the expansion of ${\cal B}_\mu$ on the
(4, 2)-dimensional supermanifold, leads to the following reduced
form of this superfield, namely; $$
\begin{array}{lcl}
{\cal B}_{\mu(r)} (x, \theta, \bar\theta) = A_\mu (x) + \theta \;
(s_{ab} A_\mu (x)) + \bar\theta\; (s_b A_\mu (x)) + \theta\;
\bar\theta\; (s_b s_{ab} A_\mu (x)).
\end{array}\eqno(4.10)
$$ The above equation demonstrates the exact derivation of the
off-shell nilpotent (anti-)BRST symmetry transformations
$s_{(a)b}$ for the $U(1)$ gauge field $A_\mu (x)$. The above
result is also same as the one derived due to the application of
the horizontality condition {\it alone}.

We concentrate on the comparison of the coefficients of the 2-form
differentials $(dx^\mu \wedge dx^\nu)$, constructed with the help
of spacetime variables alone, from the l.h.s. and r.h.s. of (4.2).
It should be noted that we shall be taking into account the
reduced form of the superfield ${\cal B}_\mu$ in our present
computation. Ultimately, we obtain the following relationship $$
\begin{array}{lcl}
&&\frac{1}{2}\; i e (dx^\mu \wedge dx^\nu)\; \Phi^* (x, \theta,
\bar\theta)  \;\bigl (\partial_\mu {\cal B}_{\nu(r)} (x, \theta,
\bar\theta) - \partial_\nu {\cal B}_{\mu(r)} (x, \theta,
\bar\theta) \bigr )\; \Phi (x, \theta, \bar\theta) \nonumber\\ &&
= \frac{1}{2}\; i e (dx^\mu \wedge dx^\nu)\; \phi^* (x) \;\bigl
(\partial_\mu  A_\nu (x) - \partial_\nu A_\mu (x) \bigr )\; \phi
(x).
\end{array}\eqno(4.11)
$$ It can be checked that $\partial_\mu {\cal B}_{\nu(r)} -
\partial_\nu {\cal B}_{\mu(r)} = \partial_\mu A_\nu - \partial_\nu
A_\mu$. Furthermore, the Abelian nature of all the fields in the
above equation allows us to cancel the gauge field part from the
l.h.s. and r.h.s. of (4.11). This entails upon the above equation
to reduce to $$
\begin{array}{lcl}
\Phi^* (x, \theta, \bar\theta) \; \Phi (x, \theta, \bar\theta) =
\phi^* (x) \; \phi (x)
\end{array}\eqno(4.12)
$$ where we have taken $e \neq 0, A_\mu \neq 0$ into
consideration. We lay emphasis on the fact that the new
relationship (4.12) is a gauge invariant condition which cannot be
obtained from the application of the horizontality condition
alone. The substitution of the expansions for the matter
superfields (cf. (4.1)) leads to the following form for the l.h.s
$$
\begin{array}{lcl}
&&\phi^*\; \phi + i \;\theta \;(\phi^* \bar f_1 + \bar f_2^*
\phi)\; +\; i \;\bar\theta \;(\phi^* f_1 + f_2^* \phi )
\nonumber\\ && + \;i\; \theta \;\bar\theta \;(\phi^* b + b^* \phi
+ i f_2^* \bar f_1 - i \bar f_2^* f_1).
\end{array}\eqno(4.13)
$$ Equating the above expressions with the r.h.s. of (4.12) leads
to the following conditions $$
\begin{array}{lcl}
&& \phi^*\; \bar f_1 + \bar f_2^*\; \phi = 0\; \qquad \;\phi^*
\;f_1 + f_2^* \;\phi = 0 \nonumber\\&& \phi^* \;b + b^* \;\phi +
i\; f_2^*\; \bar f_1 - i\; \bar f_2^* \;f_1 = 0.
\end{array}\eqno(4.14)
$$ It will be noted that, in the above, the coefficients of the
$\theta, \bar\theta$ and $\theta \bar\theta$-directions of the
above expansions, have been set equal to zero separately and
independently. At this juncture, our knowledge of the previous
section comes to our help. The following interesting choices $$
\begin{array}{lcl}
&&f_1 = - e C \phi\; \qquad \;\bar f_1 = - e \bar C \phi\;
\qquad\; f_2^* = + e \phi^* C\; \qquad \;\bar f_2^* = + e \phi^*
\bar C \nonumber\\ &&  b = - i\; e\;(B + e\; \bar C \;C)\; \phi
\;\qquad \;b^* = + i\; e\; \phi^* \;(B + e\; C\; \bar C)
\end{array}\eqno(4.15)
$$ satisfy all the above conditions quoted in (4.14). The logical
arguments in deducing the above solutions are same as in  the
previous section. The insertions of these values into the super
expansion (4.1) for the matter superfields leads to the following
reduced form $$
\begin{array}{lcl}
\Phi^{(r)} (x, \theta, \bar \theta) &=& \phi (x) + \theta\;
(s_{ab} \phi (x)) + \bar\theta\; (s_b \phi (x)) + \theta\;
\bar\theta\; (s_b s_{ab} \phi (x)) \nonumber\\ \Phi^{*(r)} (x,
\theta, \bar \theta) &=& \phi^* (x) + \theta\; (s_{ab} \phi^* (x))
+ \bar\theta\; (s_b \phi^* (x)) + \theta\; \bar\theta\; (s_b
s_{ab} \phi^* (x)).
\end{array}\eqno(4.16)
$$ The above equation demonstrates explicitly the derivation of
the off-shell nilpotent and anticommuting (anti-)BRST symmetry
transformations $s_{(a)b}$ for the matter fields of the theory in
the framework of the augmented superfield formulation.

Finally, the reduced form of the expansions, quoted for all the
superfields in (4.1), can be re-expressed in terms of the
off-shell nilpotent (anti-)BRST symmetry transformations of (2.4)
in the following uniform fashion for all the superfields of the
theory: $$
\begin{array}{lcl}
{\cal B}_{\mu(r)} (x, \theta, \bar\theta) &=& A_\mu (x) + \theta\;
 (s_{ab} A_\mu (x)) + \bar\theta\; (s_b A_\mu (x)) + \theta\;
 \bar\theta\; (s_b s_{ab} A_\mu (x)) \nonumber\\
{\cal F}_{(r)} (x, \theta, \bar\theta) &=& C (x) + \theta\;
(s_{ab} C (x)) + \bar\theta\; (s_b C (x)) + \theta\; \bar\theta\;
(s_b s_{ab} C (x)) \nonumber\\ \bar {\cal F}_{(r)} (x, \theta,
\bar\theta) &=& \bar C (x) + \theta\; (s_{ab} \bar C (x)) +
\bar\theta\; (s_b \bar C (x)) + \theta\; \bar\theta\; (s_b s_{ab}
\bar C (x)) \nonumber\\ \Phi^{(r)} (x, \theta, \bar \theta) &=&
\phi (x) + \theta\; (s_{ab} \phi (x)) + \bar\theta\; (s_b \phi
(x)) + \theta\; \bar\theta\; (s_b s_{ab} \phi (x)) \nonumber\\
\Phi^{*(r)} (x, \theta, \bar \theta) &=& \phi^* (x) + \theta\;
(s_{ab} \phi^* (x)) + \bar\theta\; (s_b \phi^* (x)) + \theta\;
\bar\theta\; (s_b s_{ab} \phi^* (x)).
\end{array}\eqno(4.17)
$$ It will be noted that, in the above expansion, the trivial
transformations $s_b C = 0, s_{ab} \bar C = 0$ have been taken
into account. The above form of the uniform expansion for all the
superfields leads to the geometrical as well as physical
interpretation for (i) the (anti-)BRST charges $Q_{(a)b}$ (and the
symmetry transformations ($s_{(a)b}$) they generate) as the
generators (cf. (2.7)) of translations (i.e. $
\mbox{Lim}_{\bar\theta \rightarrow 0} (\partial/\partial \theta),
 \mbox{Lim}_{\theta \rightarrow 0} (\partial/\partial \bar\theta)$)
along the Grassmannian directions of the six (4, 2)-dimensional
supermanifold, (ii) the nilpotency property of the (anti-)BRST
symmetry transformations (and corresponding generators) as a
couple of successive translations (i.e. $(\partial/\partial
\theta)^2 = 0, (\partial/\partial \bar\theta)^2 = 0$) along any
particular Grassmannian direction of the supermanifold, (iii) the
anticommutativity property $s_b s_{ab} + s_{ab} s_b = 0$ (and/or
$Q_b Q_{ab} + Q_{ab} Q_b = 0$) as a similar kind of relationship
(i.e. $(\partial/\partial \theta) (\partial/\partial\bar\theta) +
(\partial/\partial\bar\theta) (\partial/\partial \theta) = 0$)
existing between the translation generators along the $\theta$ and
$\bar\theta$-directions of the supermanifold, and (iv) the
internal (anti-)BRST symmetry transformations for the 4D ordinary
local field of a Lagrangian density as the translation of the
corresponding superfield along the Grassmannian direction(s) of
the supermanifold.\\

\noindent {\bf 5 Conclusions}\\

\noindent In our present investigation, we have provided a
generalization of the celebrated horizontality condition of the
usual superfield approach to BRST formalism [1-7]. This has been
done primarily for a couple of reasons. First, as is well-known,
the horizontality condition on a specifically chosen supermanifold
is {\it not} a gauge (i.e. BRST) invariant restriction. Rather, it
is intrinsically a gauge covariant restriction\footnote{For an
Abelian gauge theory, this restriction becomes a gauge invariant
restriction. In general, this condition is a gauge covariant
restriction on the gauge superfield (defined on a suitably chosen
supermanifold) and, therefore, it is not a BRST invariant
restriction on the above superfield.} because the curvature tensor
of a non-Abelian gauge theory transforms covariantly under the
SU(N) gauge transformation (which is also reflected in the
corresponding BRST transformation on it). A physical quantity,
however, has to be a gauge (i.e. BRST) invariant quantity. This is
why, in our present endeavour, we have chosen the gauge invariant
restrictions (cf. (3.2), (3.16), (4.2), (A.1)) on the matter
superfields of the appropriately chosen supermanifolds. Second,
the horizontality condition does not shed any light on the
(anti-)BRST symmetry transformations associated with the matter
fields of a given interacting gauge theory where there is an
explicit coupling between the gauge field and  matter fields.
However, in our present attempt, we have chosen restrictions on
the matter superfields of the suitably chosen supermanifolds in
such a way that they enable us to determine the exact nilpotent
(anti-)BRST symmetry transformations for the matter fields. Thus,
our present and earlier attempts [18,19] do provide a theoretical
basis for the generalization of the horizontality condition of the
usual superfield approach to BRST formalism.

One of the key features of our gauge (i.e. BRST) invariant
restrictions is the fact that they owe their origin to a pair of
(super) covariant derivatives that operate on the matter (super)
fields in unison. This specific unity of the two (super) covariant
derivatives has intimate connection with the (super) curvature
tensors of a given gauge theory. This is precisely the reason that
the geometrical interpretations of the nilpotent (anti-)BRST
symmetry transformations and their corresponding nilpotent
generators, that emerge due to the application of the
horizontality condition alone, remain intact under the gauge
invariant restrictions of our present endeavour. Furthermore, it
is very interesting to note that a single restriction on the
matter superfields of the supermanifold allows us to obtain {\it
all} the nilpotent symmetry transformations for {\it all} the
fields of a given interacting gauge theory. Thus, our present
generalization of the horizontality condition (owing its origin to
the restriction on the super gauge fields alone) is very logical,
economical and physically appealing. Our model being an
interacting Abelian gauge theory, the (super) covariant
derivatives are defined {\it only} on the matter (super) fields.
The above (super) covariant derivatives do not exist for the
Abelian (super) gauge as well as (super) (anti-)ghost fields.

In our present investigation, we have concentrated on the field
theoretical model of the 4D interacting $U(1)$ gauge theory where
there is an explicit coupling between the gauge field and complex
scalar fields (and the gauge field itself). This model is
interesting by itself because it allows the inclusion of a
renormalizable quartic potential that is found to be gauge (i.e.
BRST) invariant. This kind of potential cannot be included in an
interacting 4D (non-)Abelian gauge theory with the fermionic Dirac
fields. Furthermore, this field theoretic model allows discussions
connected with the spontaneous symmetry breaking, Goldstone
theorem, Higgs mechanism, etc., which are very useful in the
context of the gauge theory of the standard model of electro-weak
unification (see, e.g., [22-24]). Thus, to put our earlier ideas
[18,19] on a firmer footing, it is essential to check the validity
of those propositions in the context of our present field
theoretical model. It would be very interesting future endeavour
to apply the ideas of our present work and that of [18,19] to the
case of gravitational theories which resemble very much with the
non-Abelian gauge theories [22]. In fact, the idea of
horizontality condition has already been applied to gravitational
theories by Delbourgo, Jarvis and Thompson (see, e.g., [4]). This
issue is presently under investigation under our augmented
superfield formalism and our results would be reported in our
future publications [25].\\

\begin{center}

{\bf Appendix A}
\end{center}

\noindent To clarify the claims made in the footnotes before the
equations (3.2) and (3.16), we discuss here, in a concise fashion,
the derivation of the off-shell nilpotent symmetry transformations
from an alternate version of the restriction (4.2) imposed on the
matter superfields of the (4, 2)-dimensional supermanifold. This
gauge (i.e. BRST) invariant restriction is$$
\begin{array}{lcl}
\Phi (x, \theta, \bar\theta) \;\tilde {\bar {\cal D}}\; \tilde
{\bar {\cal D}}\; \Phi^* (x,\theta,\bar\theta) = \phi (x) \; \bar
D \; \bar D\; \phi^* (x)
\end{array} \eqno(A.1)
$$ where $\tilde {\bar {\cal D}} = dx^\mu (\partial_\mu - i e
{\cal B}_\mu) + d \theta (\partial_\theta - i e \bar {\cal F}) +
d\bar\theta (\partial_{\bar\theta} - i e {\cal F})$. It is evident
that the r.h.s. of the above gauge invariant condition is: $-
\frac{1}{2}\;i e \;(dx^\mu \wedge dx^\nu)\;\phi (x)\;
(\partial_\mu A_\nu - \partial_\nu A_\mu)\; \phi^* (x)$. To
compare this with the l.h.s, it is essential to expand the l.h.s.
in explicit form as given below $$
\begin{array}{lcl}
&& (dx^\mu \wedge dx^\nu)\; \Phi^* \;\bigl (\partial_\mu - i e
{\cal B}_\mu) (\partial_\nu - i e {\cal B}_\nu \bigr )\; \Phi
\nonumber\\ && - (d\theta \wedge d\theta)\; \Phi^* \;\bigl
(\partial_\theta - i e \bar {\cal F} \bigr )\;\bigl
(\partial_\theta - i e \bar {\cal F} \bigr ) \; \Phi \nonumber\\
&& - (d\bar\theta \wedge d\bar\theta)\; \Phi^* \;\bigl
(\partial_{\bar\theta} - i e  {\cal F} \bigr )\;\bigl
(\partial_{\bar\theta} - i e  {\cal F} \bigr ) \; \Phi \nonumber\\
&& - (d\theta \wedge d\bar\theta)\; \Phi^* \;\bigl [
(\partial_{\bar\theta} - i e  {\cal F}) (\partial_\theta - i e
\bar {\cal F}) +  (\partial_\theta - i e \bar {\cal F})
(\partial_{\bar\theta} - i e {\cal F}) \bigr ] \; \Phi \nonumber\\
&& + (dx^\mu \wedge d\theta)\; \Phi^*\; \bigl [ \;(\partial_\mu -
i e {\cal B}_\mu) (\partial_\theta - i e \bar {\cal F}) -
(\partial_\theta - i e \bar {\cal F}) (\partial_\mu - i e {\cal
B}_\mu) \bigr ]\; \Phi \nonumber\\ && + (dx^\mu \wedge
d\bar\theta)\; \Phi^*\; \bigl [ \;(\partial_\mu - i e {\cal
B}_\mu) (\partial_{\bar\theta} - i e  {\cal F}) -
(\partial_{\bar\theta} - i e  {\cal F}) (\partial_\mu - i e {\cal
B}_\mu) \bigr ]\; \Phi.
\end{array} \eqno(A.2)
$$ Setting, first of all, the coefficients of $d\theta \wedge
d\theta), (d\bar \theta \wedge d\bar\theta)$ and $(d\theta \wedge
d\bar\theta)$ equal to zero, we obtain the following conditions $$
\begin{array}{lcl}
&&\partial_\theta \bar {\cal F} = 0 \;\;\Rightarrow\;\; \bar B_2
(x) = 0\; \qquad \bar s (x) = 0 \nonumber\\ &&
\partial_{\bar\theta} {\cal F} = 0 \;\; \Rightarrow \;\; B_1 (x) =
0\; \qquad s (x) = 0 \nonumber\\ && \partial_\theta {\cal F} +
\partial_{\bar\theta} \bar {\cal F} = 0 \;\;\Rightarrow \;\;
\bar B_1 (x) + B_2 (x) = 0
\end{array}\eqno(A.3)
$$ when $e \neq 0, \Phi \neq 0, \Phi^* \neq 0$. Thus, according to
our earlier choice, if $B_2 (x) = B (x)$ then $\bar B_1 (x) = -
B(x)$. Insertions of these values into the expansions in (4.1)
reduces the fermionic superfields ${\cal F}$ and $\bar {\cal F}$
to ${\cal F}_{(r)}$ and $\bar {\cal F}_{(r)}$. Similarly, setting
the coefficients of the 2-form differentials $(dx^\mu \wedge
d\theta)$ and $(dx^\mu \wedge d\bar\theta)$ equal to zero leads to
the following conditions for $ e\neq 0, \Phi \neq 0, \Phi^* \neq
0$, namely; $$
\begin{array}{lcl}
\partial_\mu \bar {\cal F}_{(r)} = \partial_\theta {\cal B}_\mu\; \qquad
\partial_\mu {\cal F}_{(r)} = \partial_{\bar\theta} {\cal B}_\mu.
\end{array}\eqno(A.4)
$$ In the above, we substitute the reduced forms of the fermionic
superfields ${\cal F}$ and $\bar {\cal F}$ which are exactly same
as the ones listed in (4.7). The resulting relations, that are
found between the secondary fields of the expansion for ${\cal
B}_\mu$ superfield and the basic fields (as well as the auxiliary
field) are exactly same as the ones given in (4.9).

Finally, we compare the coefficients of the 2-form differentials
$(dx^\mu \wedge dx^\nu)$ that emerge from the l.h.s. and r.h.s. of
the restriction (A.1). In its explicit form, this equality is  $$
\begin{array}{lcl}
&& - \frac{1}{2}\; i e (dx^\mu \wedge dx^\nu)\; \Phi^* (x, \theta,
\bar\theta)  \;\bigl (\partial_\mu {\cal B}_{\nu(r)} (x, \theta,
\bar\theta) - \partial_\nu {\cal B}_{\mu(r)} (x, \theta,
\bar\theta) \bigr )\; \Phi (x, \theta, \bar\theta) \nonumber\\ &&
= - \frac{1}{2}\; i e (dx^\mu \wedge dx^\nu)\; \phi^* (x) \;\bigl
(\partial_\mu  A_\nu (x) - \partial_\nu A_\mu (x) \bigr )\; \phi
(x)
\end{array}\eqno(A.5)
$$ where ${\cal B}_{\mu(r)}$ is the reduced form of the bosonic
superfield ${\cal B}_\mu$ of the expansion (4.1) where $R_\mu =
\partial_\mu C, \bar R_\mu = \partial_\mu \bar C$ and $S_\mu =
\partial_\mu B$ have been substituted. One obtains, ultimately,
the same relationship between the matter superfields and ordinary
matter fields as given in (4.12). After this, all the steps of
computation are same as the ones given in the equations from
(4.13) till (4.17). This establishes the fact that the alternative
gauge (i.e. BRST) invariant conditions, that are mentioned in the
footnotes before equations (3.2) and (3.16), are equally useful in
obtaining the nilpotent (anti-)BRST symmetry transformations for
all the fields of the interacting $U(1)$ gauge theory where there
is an explicit coupling between $A_\mu$ field and matter fields
$\phi$ and $\phi^*$. In fact, we conclude, after some
observations, that the algebraic computations of all the steps,
for the alternative versions of the gauge (i.e. BRST) invariant
restrictions, are exactly same as the ones given in the body of
our present text except that one has to replace $e$ by $- e$ (i.e.
$ e \to -e$) in all the relevant equations.\\

\noindent {\bf Acknowledgements}\\

\noindent
 Very useful suggestions and critical comments by our
esteemed referees as well as the adjudicator are gratefully
acknowledged.

\end{document}